\documentclass[aps,prl,twocolumn,superscriptaddress,showpacs,
floatfix,nofootinbib]{revtex4-1}

\usepackage{dcolumn}
\usepackage{bm}
\usepackage{color}
\usepackage{amssymb}
\usepackage{amsmath}
\usepackage{graphicx}
\usepackage{pstricks}
\usepackage{float}
\usepackage{natbib,hyperref}

\begin{document}

\title{Charge form factor and sum rules of electromagnetic response functions in $^{12}$C}

\author{A. Lovato}
\affiliation{Argonne Leadership Computing Facility, Argonne National Laboratory, Argonne, IL 60439}
\affiliation{Physics Division, Argonne National Laboratory, Argonne, IL 60439}

\author{S. Gandolfi}
\affiliation{Theoretical Division, Los Alamos National Laboratory, Los Alamos, NM 87545}

\author{Ralph Butler}
\affiliation{Computer Science Department,Middle Tennessee State University, Murfreesboro, TN 37132}

\author{J. Carlson}
\affiliation{Theoretical Division, Los Alamos National Laboratory, Los Alamos, NM 87545}

\author{Ewing Lusk}
\affiliation{Mathematics and Computer Science Division, Argonne National Laboratory, Argonne, IL 60439}

\author{Steven C. Pieper}
\affiliation{Physics Division, Argonne National Laboratory, Argonne, IL 60439}

\author{R. Schiavilla}
\affiliation{Department of Physics, Old Dominion University, Norfolk, VA 23529}
\affiliation{Jefferson Lab, Newport News, VA 23606}

\date{\today}

\begin{abstract}
An {\it ab initio} calculation of the $^{12}$C elastic form factor,
and sum rules of longitudinal and transverse response functions
measured in inclusive $(e,e^\prime)$ scattering, is reported, based
on realistic nuclear potentials and electromagnetic currents.
The longitudinal elastic form factor and sum rule are found to be in
satisfactory agreement with available experimental data.  A direct
comparison between theory and experiment is difficult for the
transverse sum rule.  However, it is shown that the calculated one
has large contributions from two-body currents, indicating
that these mechanisms lead to a significant enhancement of
the quasi-elastic transverse response.  This fact may have implications
for the anomaly observed in recent neutrino quasi-elastic charge-changing
scattering data off $^{12}$C.
\end{abstract}


\maketitle
The current picture of the nucleus as a system of
protons and neutrons interacting among themselves via
two- and three-body forces and with external electroweak
probes via one- and two-body currents---a dynamical
framework we will refer to below as the standard nuclear
physics approach (SNPA)---has been shown to reproduce
satisfactorily a variety of empirical properties of light nuclei
with mass number $A \leq 12$, including energy 
spectra~\cite{Pieper:2001,Pieper:2008,Maris:2013,Barrett:2013,Roth:2011,Jansen:2011,Epelbaum:2010},
static properties~\cite{Pieper:2001,Marcucci:2008,Barrett:2013,Pastore:2013,Maris:2013} of low-lying states,
such as charge radii, and magnetic and quadrupole moments,
and longitudinal electron scattering~\cite{Bacca:2009,Bacca:2009prc}.
However, it has yet to be established conclusively whether such
a picture quantitatively and successfully accounts for the observed
electroweak structure and response of these systems, at least
those with $A>4$, in a wide range of energy and momentum
transfers.  This issue has acquired new and pressing relevance in
view of the anomaly seen in recent neutrino quasi-elastic charge-changing
scattering data on $^{12}$C~\cite{Aguilar:2008}, i.e., the excess, at relatively
low energy, of measured cross section relative to theoretical
calculations.  Analyses based on these calculations have led
to speculations that our present understanding of the nuclear
response to charge-changing weak probes may be
incomplete~\cite{Benhar:2010}, and, in particular, that the momentum-transfer 
dependence of the axial form factor of the nucleon
may be quite different from that obtained from analyses of pion
electro-production data~\cite{Amaldi:1979} and measurements of neutrino
and anti-neutrino reactions on protons and 
deuterons~\cite{Baker:1981,Miller:1982,Kitagaki:1983,Ahrens:1987}.
However, it should be emphasized that the calculations on which
these analyses are based use rather crude models of nuclear
structure---Fermi gas or local density approximations of the
nuclear matter spectral function---as well as simplistic treatments
of the reaction mechanism, and do not fit the picture outlined
above.  Conclusions based on them should therefore be viewed
with caution.

The present work provides the first step towards a comprehensive
study, within the SNPA, of the quasi-elastic electroweak response
functions of light nuclei.  We report an exact quantum Monte
Carlo (QMC) calculation of the elastic form factor and sum rules associated with the
longitudinal and transverse response functions measured in
inclusive electron scattering experiments on $^{12}$C.  These sum
rules are defined as~\cite{Carlson:2002}
\begin{equation}
\label{eq:sr}
S_\alpha (q)= C_\alpha \int_{\omega^+_{\rm th}}^\infty 
{\rm d}\omega\, \frac{R_\alpha(q,\omega)} {G_E^{p\, 2}(Q^2)}\ ,
\end{equation}
where $R_\alpha(q,\omega)$ is the longitudinal ($\alpha=L$) 
or transverse ($\alpha=T$) response function, $q$ and $\omega$
are the momentum and energy transfers, $\omega_{\rm th}$ is the
energy transfer corresponding to the inelastic threshold (the first
excited-state energy is at 4.44 MeV relative to the ground state in
$^{12}$C), $G_E^p(Q^2)$ is the proton electric form factor
evaluated at four-momentum transfer $Q^2=q^2-\omega^2$, and
the $C_\alpha$'s are appropriate normalization factors, given by
\begin{equation}
C_L=\frac{1}{Z} \ , 
\qquad C_T=\frac{2}{\left(Z\, \mu_p^2+N\, \mu_n^2\right)}\, \frac{m^2}{q^2} \ .
\end{equation}
Here $m$ is the nucleon mass, and $Z$ ($N$) and $\mu_p$ ($\mu_n$) are
the proton (neutron) number and magnetic moment, respectively.  These
factors have been introduced so that $S_\alpha( q\rightarrow \infty )\simeq 1$
under the approximation that the nuclear charge and current operators
originate solely from the charge and spin magnetization of individual
protons and neutrons and that relativistic corrections to these one-body
operators---such as the Darwin-Foldy and spin-orbit terms in the charge
operator---are ignored.

It is well known~\cite{McVoy:1962} that the sum rules above can be expressed
as ground-state expectation values of the type
\begin{equation}
\label{eq:sri}
S_\alpha(q)\!=\!C_\alpha  \Big[ \langle 0|
O_\alpha^\dagger({\bf q}) \, O_\alpha({\bf q}) |0\rangle -|\langle 0;{\bf q}|
 O_\alpha({\bf q}) |0 \rangle|^2 \Big]  \ ,
\end{equation}
where $O_\alpha({\bf q})$ is either the charge $\rho({\bf q})$ ($\alpha=L$)
or transverse current ${\bf j}_\perp({\bf q})$ ($\alpha=T$) operator divided
by $G_E^p(Q^2)$, $|0;{\bf q}\rangle$ denotes the ground state of the nucleus
recoiling with total momentum ${\bf q}$, and averages over the spin projections
have been suppressed
because $^{12}$C has $J^\pi$=$\, 0^+$.  
The $S_\alpha(q)$ as defined in
Eq.~(\ref{eq:sr}) only includes the inelastic contribution to $R_\alpha(q,\omega)$,
i.e., the elastic contribution represented by the second term on the r.h.s.~of
Eq.~(\ref{eq:sri}) has been removed.  It is proportional to the longitudinal
($F_L$) or transverse ($F_T$) elastic form factor.  For $^{12}$C, $F_T$
vanishes, while $F_L(q)$ (to be discussed below) is given by
$F_L(q)\!=\!G_E^p(Q^2_{\rm el}) \,\langle 0;{\bf q}|  O_L({\bf q}) |0 \rangle/Z$,
with the four-momentum transfer $Q^2_{\rm el}=q^2-\omega_{\rm el}^2$
and $\omega_{\rm el}$ corresponding to elastic scattering,
$\omega_{\rm el}=\sqrt{q^2+m_A^2}-m_A$ ($m_A$ is the $^{12}$C mass). 
  
The sum rules $S_\alpha(q)$ provide a useful tool for studying
integral properties of the response of the nucleus to an external
electromagnetic probe, and their calculation does not
require any knowledge of the complicated structure of the nuclear
excitation spectrum.  Unfortunately, direct comparison between
the calculated and experimentally extracted sum rules cannot
be made unambiguously for two reasons.  First, the experimental
determination of $S_\alpha$ requires measuring the associated
$R_\alpha$ in the whole energy-transfer region, from threshold up
to $\infty$.  Inclusive electron scattering experiments only allow
access to the space-like region of the four-momentum transfer
($ \omega < q$).  While the response in the time-like region ($\omega > q$)
could, in principle, be measured via $e^+ e^-$ annihilation, no
such experiments have been carried out to date.  Therefore, for a
meaningful comparison between theory and experiment, one needs 
to estimate the strength outside the region covered by the experiment. 
We will return to this issue below.  For the moment, it suffices to say 
that in the past this has been accomplished in the case of $S_L(q)$ either
by extrapolating the data~\cite{Jourdan:1996} or, in the few-nucleon
systems, by parametrizing the high-energy tail and using energy-weighted
sum rules to constrain it~\cite{Schiavilla:1989,Schiavilla:1993}.

The second reason that direct comparison of
theoretical and ``experimental'' sum rules is difficult lies in the inherent
inadequacy of the current SNPA to account for explicit pion production
mechanisms.  The latter mostly affect the transverse response and
make its $\Delta$-peak region outside the range of applicability of this
approach.  However, the one- and two-body charge and current
operators adopted in the present work should provide a realistic
and quantitative description of  both longitudinal and transverse
response in the quasi-elastic region, where nucleon and (virtual)
pion degrees of freedom are expected to be dominant.  At low
and intermediate momentum transfers ($q\lesssim 400$ MeV/c),
the quasi-elastic and $\Delta$-peak are well separated, and 
it is therefore reasonable to study sum rules of the transverse
response. 

The ground-state wave function of $^{12}$C is obtained from a 
Green's function Monte Carlo (GFMC) solution of the Schr\"odinger equation
including the Argonne $v_{18}$ (AV18) two-nucleon ($N\!N$)~\cite{Wiringa:1995}
and Illinois-7 (IL7) three-nucleon ($N\!N\!N$)~\cite{Pieper:2008} potentials.
The AV18 consists of a long-range component induced by one-pion
exchange (OPE) and intermediate-to-short range components modeled
phenomenologically,  and fits the $N\!N$ scattering database for energies
up to $E_{lab}=350$ MeV with a $\chi^2$ per datum close to one.
The IL7 includes a central (albeit isospin dependent) short-range repulsive
term and two- and three-pion-exchange mechanisms involving excitation
of intermediate $\Delta$ resonances.  Its strength is determined by four
parameters which are fixed by a best fit to the energies of 17 low-lying
states of nuclei in the mass range $A \leq 10$, obtained in combination with
the AV18  $N\!N$ potential.  As already noted, the AV18+IL7 Hamiltonian
reproduces well the spectra of nuclei with $A\leq10$~\cite{Pieper:2008}---in
particular, the attraction provided by the Illinois $N\!N\!N$ potentials in isospin
3/2 triplets is crucial for the $p$-shell nuclei---and the $p$-wave resonances
with $J^\pi=(3/2)^-$ and $(1/2)^-$ in low-energy neutron scattering off
$^4$He~\cite{Nollett:2007}.

The $^{12}$C ground state wave function is evolved in imaginary time by GFMC from a
variational (VMC) wave function that contains both explicit alpha clustering and the
five possible $J^\pi$=$0^+$ $p$-shell states.  These are multiplied by two- and
three-body non-central correlations~\cite{Pudliner:1997}.  Our ground-state energy and RMS
charge radius are \hbox{--93.3(4)}~MeV and 2.46(2)~fm, respectively, in good agreement with
the experimental values of \hbox{--92.16}~MeV and $(2.471 \pm 0.005)$ fm~\cite{Sick:1982}.

Realistic models for the electromagnetic charge and current operators include
one- and two-body terms (see Ref.~\cite{Shen:2012} for a recent overview).  The
former follow from a non-relativistic expansion of the single-nucleon four-current,
in which corrections proportional to $1/m^2$ are retained.  Leading two-body
terms are derived from the static part of the $N\!N$ potential (the AV18 in the
present case), which is assumed to be due to exchanges of effective
pseudo-scalar ($\pi$-like) and vector ($\rho$-like) mesons.  The corresponding
charge and current operators are constructed from non-relativistic reductions
of Feynman amplitudes with the $\pi$-like and $\rho$-like effective propagators
projected out of the central, spin-spin and tensor components of the $N\!N$
potential.  They contain no free parameters, and their short-range
behavior is consistent with that of the potential.  In particular, the longitudinal
part of these two-body currents satisfies, by construction, current conservation
with the (static part of the) $N\!N$ potential.  Additional contributions---purely
transverse and hence unconstrained by current conservation---come from
$M1$-excitation of $\Delta$ resonances treated perturbatively in the intermediate state (for the current)
and from the $\rho\pi\gamma$ transition mechanism (for the charge and current).
For these, the values of the various coupling constants are taken from experiment~\cite{Shen:2012}.
As documented in Refs.~\cite{Carlson:2002,Carlson:1998,Marcucci:2005}, these charge and current
operators reproduce quite well a variety of few-nucleon electromagnetic observables,
ranging from elastic form factors to low-energy radiative capture cross sections to
the quasi-elastic response in inclusive $(e, e^\prime)$ scattering at intermediate energies.

The spin-orbit and convection terms in $O_L({\bf q})$ and $O_T({\bf q})$ require
gradients of both the bra and ket
in Eq.~(\ref{eq:sri}); however, we cannot compute gradients of the evolved GFMC wave
function.  Therefore we compute these terms for only the VMC
wave function and add them perturbatively to the GFMC results.  They are
generally quite small, although the convection term is significant for small $q$,
see Fig.~\ref{fig:f3} below.

The calculations were made on Argonne's IBM Blue Gene/Q (Mira). Our GFMC program
uses the Asynchronous Dynamic Load Balancing (ADLB) library~\cite{Lusk:2010} to
achieve parallelization to more than 250,000 MPI processes with 80\% efficiency
to calculate the energy.  The
calculations of operators presented here require much more memory than just the
energy evaluation and we typically used four MPI processes on each 16 Gbyte node.
We achieve good OpenMP scaling in each process:
using 16 threads (the most possible) instead of only 4 reduces the time
per configuration per $q$-value from about 12 to 6 minutes.
For each Monte Carlo configuration, we averaged over 12 directions of $\hat{\bf q}$
in Eq.~(\ref{eq:sri}); these were in four groups of three orthogonal directions
obtained by implementing the method of uniformly distributed random rotations on a 
unit sphere~\cite{Arvo:1992}. The 12 calculations for each of 21 magnitudes of $q$
(252 independent calculations) were distributed to different
MPI processes by ADLB, with an efficiency above 95\% on more than 32,000 MPI
processes. 

\begin{figure}[bth]
\includegraphics[width=8cm,height=6cm]{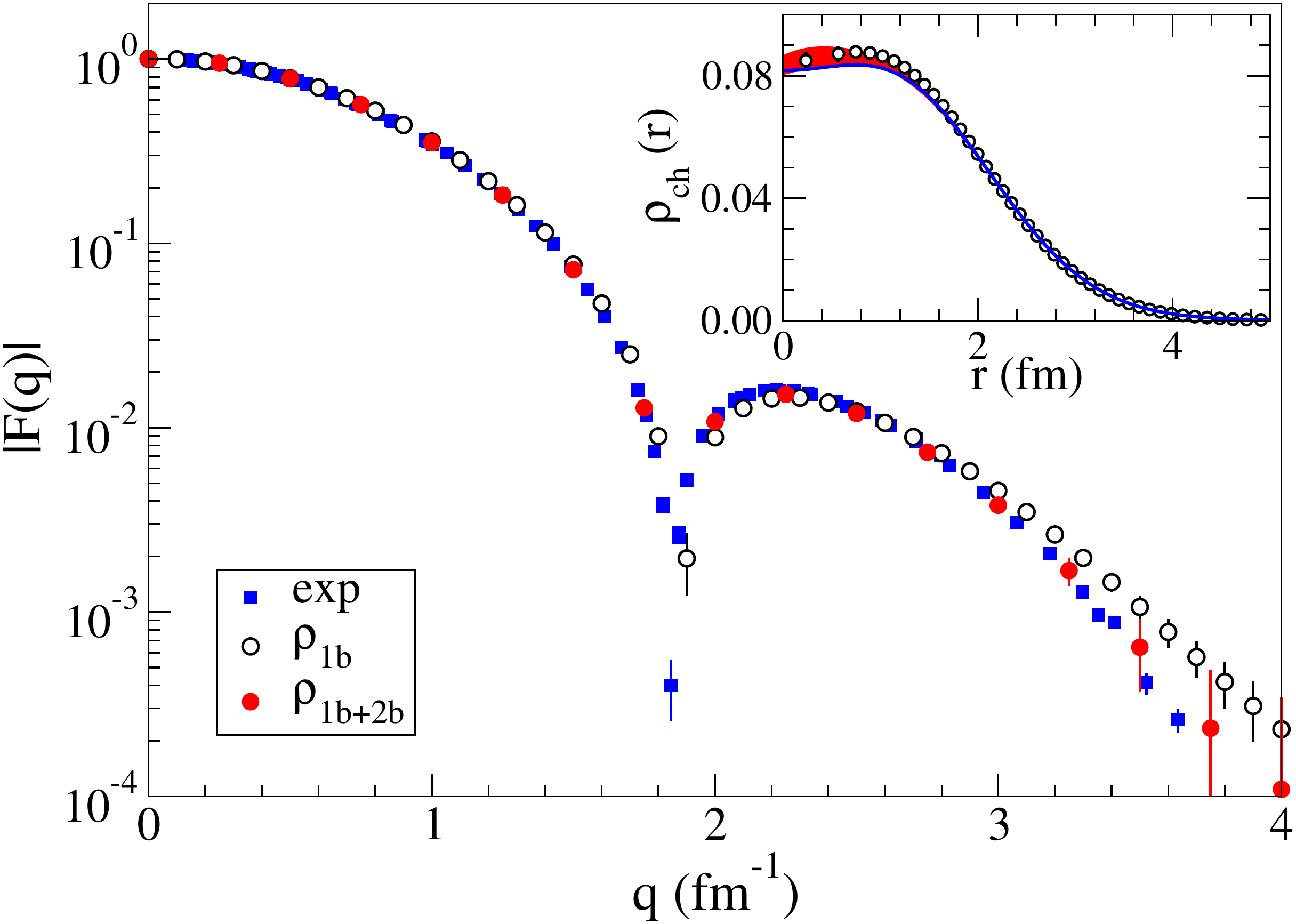}
\caption{(Color online)  The longitudinal elastic form factor of $^{12}$C
obtained from the AV18+IL7 Hamiltonian with one-body only (empty circles) and
one- and two-body (solid circles) terms in the charge operator is compared to experimental
data.  Also displayed are the statistical errors of the QMC calculation.
The inset shows the calculated charge density in coordinate space with one-body (empty circles) 
and (one+two)-body (red band) terms compared with
an analysis of the experimental data (solid line)~\cite{deVries:1987}.}
\label{fig:f1}
\end{figure}

The calculated longitudinal elastic form factor ($F_L$) of $^{12}$C
is compared to experimental data in Fig.~\ref{fig:f1}.  These data are
from an unpublished compilation by Sick~\cite{Sick:2013,Sick:1982}, and are well
reproduced by theory over the whole range of momentum transfers.  The results
labeled one-body (1b) include, in addition to the proton, the neutron contribution as
well as the Darwin-Foldy and spin-orbit relativistic corrections to the single-nucleon
charge operator, while those labeled two-body (2b) also contain the contributions due
to the $\pi$-like, $\rho$-like, and $\rho\pi\gamma$ (two-body) charge operators.
These two-body contributions are negligible at low $q$, and become
appreciable only for $q > 3$ fm$^{-1}$, where they interfere destructively
with the one-body contributions bringing theory into closer agreement with
experiment.  The Simon~\cite{Simon:1980}, Galster~\cite{Galster:1971}, and
H\"ohler~\cite{Hohler:1976} parametrizations are used for the proton electric, neutron
electric, and proton and neutron magnetic form factors, respectively.

\begin{figure}[bth]
\includegraphics[width=8cm,height=6cm]{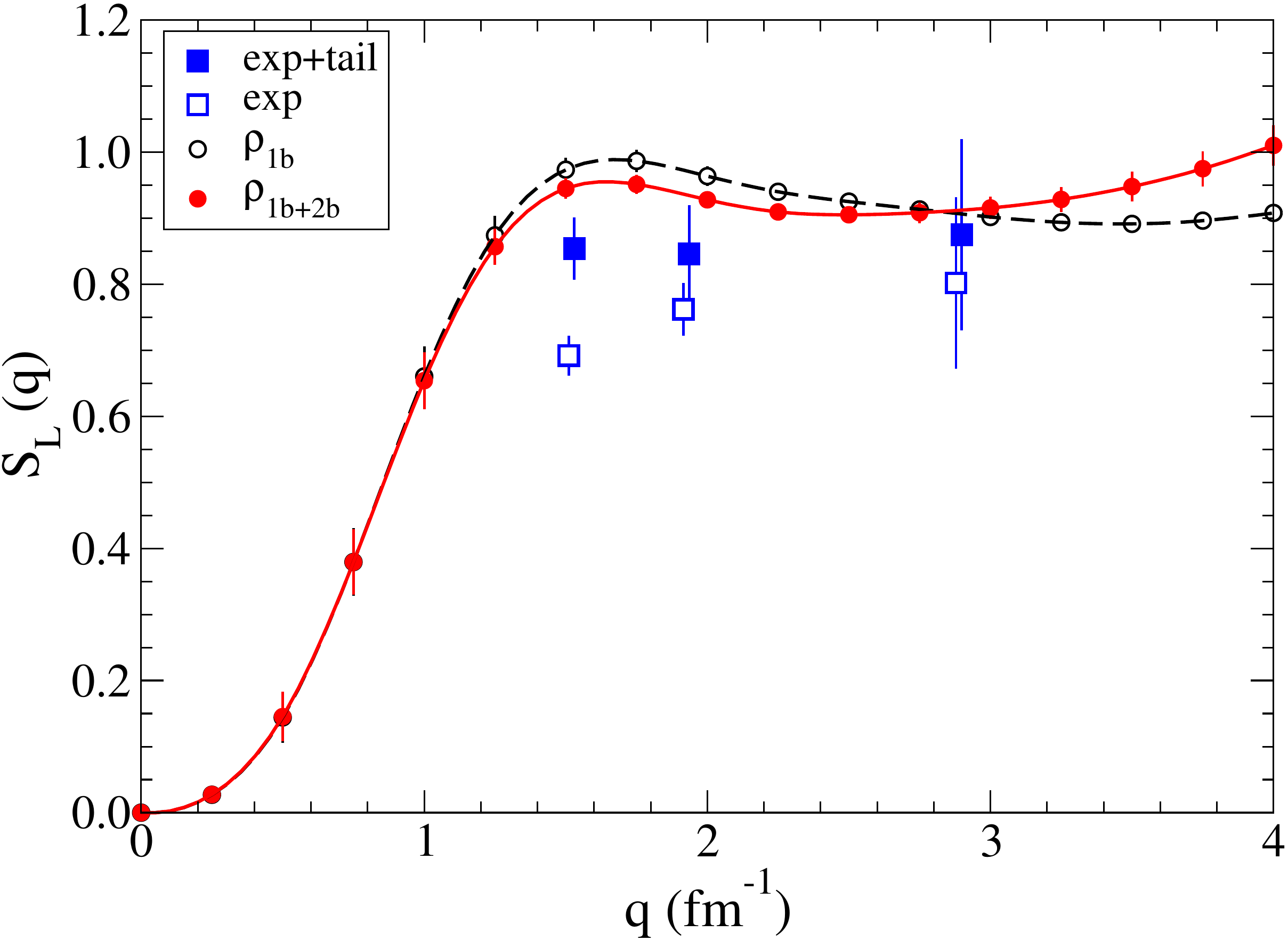}
\caption{(Color online)  The longitudinal sum rule of $^{12}$C
obtained from the AV18+IL7 Hamiltonian with one-body only (empty circles, dashed line) and
one- and two-body (solid circles, solid line) terms in the charge operator is compared to experimental
data without (empty squares), labeled exp, and with (solid squares),
labeled exp+tail, the tail contribution, see text.  Also displayed
are the statistical errors of the QMC calculation.}
\label{fig:f2}
\end{figure}

In Figs.~\ref{fig:f2} and~\ref{fig:f3}, we show by the open squares the experimental sum rules 
$S_L(q)$ and $S_T(q)$ obtained by integrating up to $\omega_{\rm max}$ the
longitudinal and transverse response functions (divided by the square
of $G^p_E$) extracted from world data on inclusive $(e,e^\prime)$ scattering
off $^{12}$C~\cite{Jourdan:1996}.  
For $q$=1.53, 1.94, and 2.90 fm$^{-1}$, $\omega_{\rm max}$
in the longitudinal (transverse) case corresponds to, respectively, 140, 210, and
345 (140, 180, 285) MeV.  We also show by the solid squares the experimental sum rules
obtained by estimating the contribution of strength in the region $\omega > \omega_{\rm max}$.
This estimate $\Delta S_\alpha (q)$ is made by assuming that for $\omega > \omega_{\rm max}$,
i.e., well beyond the quasi-elastic peak, the (longitudinal or transverse) response in a
nucleus like $^{12}$C ($R_\alpha^A$) is proportional to that in the deuteron
($R_\alpha^{\, d}$), which can be accurately calculated~\cite{Shen:2012}. In particular,
$R_\alpha^{\, d}$ has been calculated using AV18, but very similar results are obtained by using
N3LO \cite{Entem:2003} instead.   Thus, we set
$R^A_\alpha(q,\omega > \omega_{\rm max}) = \lambda(q)\, R^{\, d}_\alpha(q,\omega)$, and
determine $\lambda(q)$ by matching the experimental $^{12}$C response to the calculated
deuteron one.  In practice, $\Delta S_\alpha(q)$ follows from
\begin{equation}
\Delta S_\alpha(q)=\lambda(q)\, C_\alpha^A\left[ \frac{S^d_\alpha(q)}{C_\alpha^d} -
\int_{\omega_{\rm th}^+}^{\omega_{\rm max} } \frac{R^d_\alpha(q,\omega)}
 {G_E^{p\, 2}(Q^2)} \right] \ ,
\end{equation}
where the $C_\alpha^A$ and $C_\alpha^d$ are the normalization factors
associated with the nucleus and deuteron, respectively, and $S^d_\alpha(q)$
is the deuteron sum rule.  It is worthwhile emphasizing that, for the transverse
case, this estimate is particularly uncertain for the reasons explained earlier.
In particular, the data on $R_T$ at $q$=1.94 and 2.90 fm$^{-1}$~\cite{Jourdan:1996}
suggest that at $\omega \sim \omega_{\rm max}$ there might be already
significant strength that has leaked in from the $\Delta$-peak region.

The scaling assumption above assumes that the high-energy
part of the response is dominated by two-nucleon physics, and that the
most important contribution is from deuteron-like $np$ pairs.
The high-energy response can be obtained from the Fourier
transform of the short-time response
$(2\pi)\, S_\alpha (q,\omega) =  \int {\rm d}t \, \langle 0 | O_\alpha^\dagger ({\bf q})
\exp ( - i H t ) O_\alpha({\bf q}) | 0 \rangle$, or equivalently from the small
imaginary time-dependence of the propagator.  At short times the
full propagator is governed by the product of pair propagators (assuming
three-nucleon interactions are weak), and hence we expect the
scaling with deuteron-like pairs.

\begin{figure}[bth]
\includegraphics[width=8cm,height=6cm]{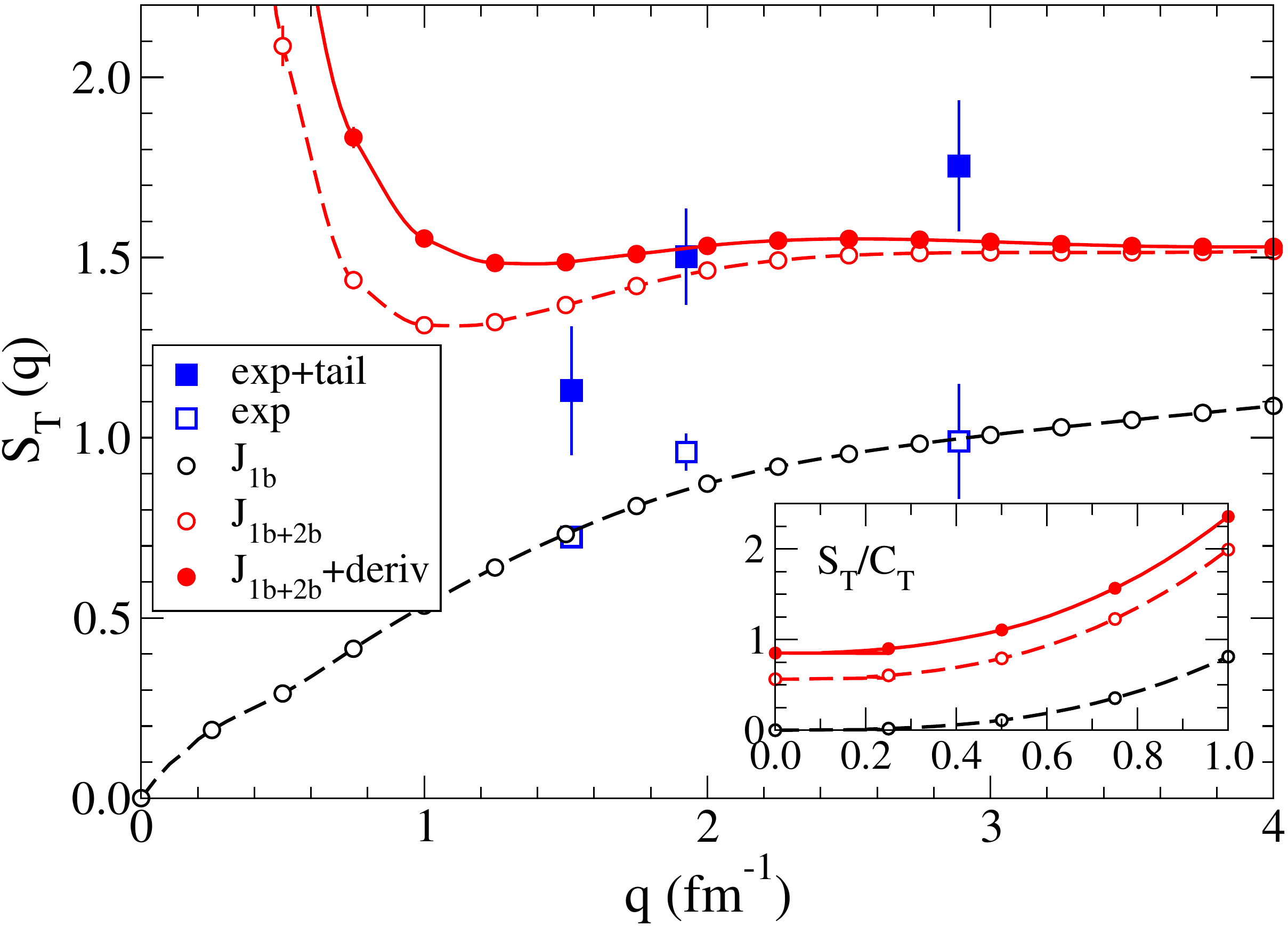}
\caption{(Color online) Same as in Fig.~\ref{fig:f2}, but for the
transverse sum rule.  The open symbols do not contain derivative terms
while a VMC evaluation of the derivative terms is included for
the solid red dots.
The inset shows $S_T(q)/C_T$ in the
small $q$-region.}
\label{fig:f3}
\end{figure}

The sum rules computed with the AV18+IL7 Hamiltonian
and one-body only or one- and two-body terms in the
charge ($S_L$) and current ($S_T$) operators are shown,
respectively, by the dashed and solid lines in Figs.~\ref{fig:f2}--\ref{fig:f3}.
In the small $q$ limit, $S_L(q)$ vanishes quadratically, while
the divergent behavior in $S_T(q)$ is due to the $1/q^2$ present
in the normalization factor $C_T$.  In this limit,
$O_T({\bf q}\!=\!0)=i\left[\, H\, , \, \sum_i {\bf r}_i \, P_i \, \right]$~\cite{Carlson:1998,Marcucci:2005},
where $H$ is the Hamiltonian and $P_i$ is the proton projector,
and therefore $S_T(q)/C_T$ is finite, indeed the associated
strength is due to collective excitations of electric-dipole
type in the nucleus.  In the large $q$ limit, the one-body
sum rules differ from one because of relativistic corrections
in $O_L({\bf q})$, primarily the Darwin-Foldy term which gives
a contribution $-\eta/(1+\eta)$ to $S_L^{\rm 1b}(q)$, where
$\eta \simeq q^2/(4\, m^2)$, and because of the convection
term in $O_T({\bf q})$, which gives a contribution $\simeq (4/3)\, C_T\, T_p/m$ to
$S^{\rm 1b}_T(q)$, where $T_p$ is the proton kinetic energy in the nucleus.

The calculated $S_L(q)$ is in satisfactory agreement with the experimental
values, including tail contributions, and no significant quenching of longitudinal
strength is observed.  Since the experimental $R_L(q,\omega)$ is divided
out by the (square of the) free proton electric form factor, one is led to
conclude that there is no evidence for in-medium modifications of the nucleon
electromagnetic form factors, as advocated, for example, by the quark-meson
coupling model of nucleon and nuclear structure~\cite{Lu:1999,Guichon:2004}.

In contrast to $S_L$, the transverse sum rule has large two-body
contributions---at $q=2.5$ fm$^{-1}$; these increase $S_T^{\rm 1b}$
by about 50\%.  Studies of Euclidean transverse response functions in the
few-nucleon systems within the same SNPA adopted here~\cite{Carlson:2002}
suggest that a significant portion of this excess transverse strength is
in the quasi-elastic region.  Clearly, a direct QMC calculation of the $^{12}$C
response functions is needed to resolve this issue conclusively.  It will
also be interesting to see the extent to which these considerations---in particular,
the major role played by two-body currents---will remain valid in the weak sector probed
in neutrino scattering, and possibly provide an explanation for the observed
$^{12}$C anomaly mentioned in the introduction.
%

We thank D.\ Day, J.\ Jourdan, and, particularly, I.\ Sick for providing
us with the data on the elastic form factor and inclusive response
functions of $^{12}$C, and for correspondence in reference to
various aspects of the analysis of these data.
We also thank S.\ Pastore and R.B.\ Wiringa for advice in the
early phases of this work.
The calculations were performed with a grant of Early-Science time on
Mira, Argonne's IBM Blue Gene/Q, and also used resources 
provided by Los Alamos Open Supercomputing and by the National Energy
Research Scientific Computing Center (NERSC).
This research is supported by the U.S.~Department of Energy, Office of
Nuclear Physics, under contracts DE-AC02-06CH11357 (A.L. and S.C.P.),
DE-AC02-05CH11231 (S.G.~and J.C.), DE-AC05-06OR23177 (R.S.), the NUCLEI SciDAC program
and by the LANL LDRD program.

\bibliography{biblio}

\begin{thebibliography}{40}%
\makeatletter
\providecommand \@ifxundefined [1]{%
 \@ifx{#1\undefined}
}%
\providecommand \@ifnum [1]{%
 \ifnum #1\expandafter \@firstoftwo
 \else \expandafter \@secondoftwo
 \fi
}%
\providecommand \@ifx [1]{%
 \ifx #1\expandafter \@firstoftwo
 \else \expandafter \@secondoftwo
 \fi
}%
\providecommand \natexlab [1]{#1}%
\providecommand \enquote  [1]{``#1''}%
\providecommand \bibnamefont  [1]{#1}%
\providecommand \bibfnamefont [1]{#1}%
\providecommand \citenamefont [1]{#1}%
\providecommand \href@noop [0]{\@secondoftwo}%
\providecommand \href [0]{\begingroup \@sanitize@url \@href}%
\providecommand \@href[1]{\@@startlink{#1}\@@href}%
\providecommand \@@href[1]{\endgroup#1\@@endlink}%
\providecommand \@sanitize@url [0]{\catcode `\\12\catcode `\$12\catcode
  `\&12\catcode `\#12\catcode `\^12\catcode `\_12\catcode `\%12\relax}%
\providecommand \@@startlink[1]{}%
\providecommand \@@endlink[0]{}%
\providecommand \url  [0]{\begingroup\@sanitize@url \@url }%
\providecommand \@url [1]{\endgroup\@href {#1}{\urlprefix }}%
\providecommand \urlprefix  [0]{URL }%
\providecommand \Eprint [0]{\href }%
\providecommand \doibase [0]{http://dx.doi.org/}%
\providecommand \selectlanguage [0]{\@gobble}%
\providecommand \bibinfo  [0]{\@secondoftwo}%
\providecommand \bibfield  [0]{\@secondoftwo}%
\providecommand \translation [1]{[#1]}%
\providecommand \BibitemOpen [0]{}%
\providecommand \bibitemStop [0]{}%
\providecommand \bibitemNoStop [0]{.\EOS\space}%
\providecommand \EOS [0]{\spacefactor3000\relax}%
\providecommand \BibitemShut  [1]{\csname bibitem#1\endcsname}%
\let\auto@bib@innerbib\@empty
\bibitem [{\citenamefont {Pieper}\ and\ \citenamefont
  {Wiringa}(2001)}]{Pieper:2001}%
  \BibitemOpen
  \bibfield  {author} {\bibinfo {author} {\bibfnamefont {S.~C.}\ \bibnamefont
  {Pieper}}\ and\ \bibinfo {author} {\bibfnamefont {R.~B.}\ \bibnamefont
  {Wiringa}},\ }\href {\doibase 10.1146/annurev.nucl.51.101701.132506}
  {\bibfield  {journal} {\bibinfo  {journal} {Ann.Rev.Nucl.Part.Sci.}\ }\textbf
  {\bibinfo {volume} {51}},\ \bibinfo {pages} {53} (\bibinfo {year} {2001})},\
  \Eprint {http://arxiv.org/abs/nucl-th/0103005} {arXiv:nucl-th/0103005
  [nucl-th]} \BibitemShut {NoStop}%
\bibitem [{\citenamefont {Pieper}(2008)}]{Pieper:2008}%
  \BibitemOpen
  \bibfield  {author} {\bibinfo {author} {\bibfnamefont {S.~C.}\ \bibnamefont
  {Pieper}},\ }\href {\doibase 10.1063/1.2932280} {\bibfield  {journal}
  {\bibinfo  {journal} {AIP Conf. Proc.}\ }\textbf {\bibinfo {volume} {1011}},\
  \bibinfo {pages} {143} (\bibinfo {year} {2008})}\BibitemShut {NoStop}%
\bibitem [{\citenamefont {Maris}\ and\ \citenamefont {Vary}()}]{Maris:2013}%
  \BibitemOpen
  \bibfield  {author} {\bibinfo {author} {\bibfnamefont {P.}~\bibnamefont
  {Maris}}\ and\ \bibinfo {author} {\bibfnamefont {J.~P.}\ \bibnamefont
  {Vary}},\ }\href@noop {} {\bibfield  {journal} {\bibinfo  {journal}
  {International Journal of Modern Physics E}\ }\textbf {\bibinfo {volume} {To
  be published}}}\BibitemShut {NoStop}%
\bibitem [{\citenamefont {Barrett}\ \emph {et~al.}(2013)\citenamefont
  {Barrett}, \citenamefont {Navr‡til},\ and\ \citenamefont
  {Vary}}]{Barrett:2013}%
  \BibitemOpen
  \bibfield  {author} {\bibinfo {author} {\bibfnamefont {B.~R.}\ \bibnamefont
  {Barrett}}, \bibinfo {author} {\bibfnamefont {P.}~\bibnamefont {Navr‡til}}, \
  and\ \bibinfo {author} {\bibfnamefont {J.~P.}\ \bibnamefont {Vary}},\ }\href
  {\doibase 10.1016/j.ppnp.2012.10.003} {\bibfield  {journal} {\bibinfo
  {journal} {Progress in Particle and Nuclear Physics}\ }\textbf {\bibinfo
  {volume} {69}},\ \bibinfo {pages} {131 } (\bibinfo {year}
  {2013})}\BibitemShut {NoStop}%
\bibitem [{\citenamefont {Roth}\ \emph {et~al.}(2011)\citenamefont {Roth},
  \citenamefont {Langhammer}, \citenamefont {Calci}, \citenamefont {Binder},\
  and\ \citenamefont {Navr\'atil}}]{Roth:2011}%
  \BibitemOpen
  \bibfield  {author} {\bibinfo {author} {\bibfnamefont {R.}~\bibnamefont
  {Roth}}, \bibinfo {author} {\bibfnamefont {J.}~\bibnamefont {Langhammer}},
  \bibinfo {author} {\bibfnamefont {A.}~\bibnamefont {Calci}}, \bibinfo
  {author} {\bibfnamefont {S.}~\bibnamefont {Binder}}, \ and\ \bibinfo {author}
  {\bibfnamefont {P.}~\bibnamefont {Navr\'atil}},\ }\href {\doibase
  10.1103/PhysRevLett.107.072501} {\bibfield  {journal} {\bibinfo  {journal}
  {Phys. Rev. Lett.}\ }\textbf {\bibinfo {volume} {107}},\ \bibinfo {pages}
  {072501} (\bibinfo {year} {2011})}\BibitemShut {NoStop}%
\bibitem [{\citenamefont {Jansen}\ \emph {et~al.}(2011)\citenamefont {Jansen},
  \citenamefont {Hjorth-Jensen}, \citenamefont {Hagen},\ and\ \citenamefont
  {Papenbrock}}]{Jansen:2011}%
  \BibitemOpen
  \bibfield  {author} {\bibinfo {author} {\bibfnamefont {G.~R.}\ \bibnamefont
  {Jansen}}, \bibinfo {author} {\bibfnamefont {M.}~\bibnamefont
  {Hjorth-Jensen}}, \bibinfo {author} {\bibfnamefont {G.}~\bibnamefont
  {Hagen}}, \ and\ \bibinfo {author} {\bibfnamefont {T.}~\bibnamefont
  {Papenbrock}},\ }\href {\doibase 10.1103/PhysRevC.83.054306} {\bibfield
  {journal} {\bibinfo  {journal} {Phys. Rev. C}\ }\textbf {\bibinfo {volume}
  {83}},\ \bibinfo {pages} {054306} (\bibinfo {year} {2011})}\BibitemShut
  {NoStop}%
\bibitem [{\citenamefont {Epelbaum}\ \emph {et~al.}(2010)\citenamefont
  {Epelbaum}, \citenamefont {Krebs}, \citenamefont {Lee},\ and\ \citenamefont
  {Mei\ss{}ner}}]{Epelbaum:2010}%
  \BibitemOpen
  \bibfield  {author} {\bibinfo {author} {\bibfnamefont {E.}~\bibnamefont
  {Epelbaum}}, \bibinfo {author} {\bibfnamefont {H.}~\bibnamefont {Krebs}},
  \bibinfo {author} {\bibfnamefont {D.}~\bibnamefont {Lee}}, \ and\ \bibinfo
  {author} {\bibfnamefont {U.-G.}\ \bibnamefont {Mei\ss{}ner}},\ }\href
  {\doibase 10.1103/PhysRevLett.104.142501} {\bibfield  {journal} {\bibinfo
  {journal} {Phys. Rev. Lett.}\ }\textbf {\bibinfo {volume} {104}},\ \bibinfo
  {pages} {142501} (\bibinfo {year} {2010})}\BibitemShut {NoStop}%
\bibitem [{\citenamefont {Marcucci}\ \emph {et~al.}(2008)\citenamefont
  {Marcucci}, \citenamefont {Pervin}, \citenamefont {Pieper}, \citenamefont
  {Schiavilla},\ and\ \citenamefont {Wiringa}}]{Marcucci:2008}%
  \BibitemOpen
  \bibfield  {author} {\bibinfo {author} {\bibfnamefont {L.~E.}\ \bibnamefont
  {Marcucci}}, \bibinfo {author} {\bibfnamefont {M.}~\bibnamefont {Pervin}},
  \bibinfo {author} {\bibfnamefont {S.~C.}\ \bibnamefont {Pieper}}, \bibinfo
  {author} {\bibfnamefont {R.}~\bibnamefont {Schiavilla}}, \ and\ \bibinfo
  {author} {\bibfnamefont {R.~B.}\ \bibnamefont {Wiringa}},\ }\href {\doibase
  10.1103/PhysRevC.78.065501} {\bibfield  {journal} {\bibinfo  {journal} {Phys.
  Rev. C}\ }\textbf {\bibinfo {volume} {78}},\ \bibinfo {pages} {065501}
  (\bibinfo {year} {2008})}\BibitemShut {NoStop}%
\bibitem [{\citenamefont {Pastore}\ \emph {et~al.}(2013)\citenamefont
  {Pastore}, \citenamefont {Pieper}, \citenamefont {Schiavilla},\ and\
  \citenamefont {Wiringa}}]{Pastore:2013}%
  \BibitemOpen
  \bibfield  {author} {\bibinfo {author} {\bibfnamefont {S.}~\bibnamefont
  {Pastore}}, \bibinfo {author} {\bibfnamefont {S.~C.}\ \bibnamefont {Pieper}},
  \bibinfo {author} {\bibfnamefont {R.}~\bibnamefont {Schiavilla}}, \ and\
  \bibinfo {author} {\bibfnamefont {R.~B.}\ \bibnamefont {Wiringa}},\ }\href
  {\doibase 10.1103/PhysRevC.87.035503} {\bibfield  {journal} {\bibinfo
  {journal} {Phys. Rev. C}\ }\textbf {\bibinfo {volume} {87}},\ \bibinfo
  {pages} {035503} (\bibinfo {year} {2013})}\BibitemShut {NoStop}%
\bibitem [{\citenamefont {Bacca}\ \emph
  {et~al.}(2009{\natexlab{a}})\citenamefont {Bacca}, \citenamefont {Barnea},
  \citenamefont {Leidemann},\ and\ \citenamefont {Orlandini}}]{Bacca:2009}%
  \BibitemOpen
  \bibfield  {author} {\bibinfo {author} {\bibfnamefont {S.}~\bibnamefont
  {Bacca}}, \bibinfo {author} {\bibfnamefont {N.}~\bibnamefont {Barnea}},
  \bibinfo {author} {\bibfnamefont {W.}~\bibnamefont {Leidemann}}, \ and\
  \bibinfo {author} {\bibfnamefont {G.}~\bibnamefont {Orlandini}},\ }\href
  {\doibase 10.1103/PhysRevLett.102.162501} {\bibfield  {journal} {\bibinfo
  {journal} {Phys. Rev. Lett.}\ }\textbf {\bibinfo {volume} {102}},\ \bibinfo
  {pages} {162501} (\bibinfo {year} {2009}{\natexlab{a}})}\BibitemShut
  {NoStop}%
\bibitem [{\citenamefont {Bacca}\ \emph
  {et~al.}(2009{\natexlab{b}})\citenamefont {Bacca}, \citenamefont {Barnea},
  \citenamefont {Leidemann},\ and\ \citenamefont {Orlandini}}]{Bacca:2009prc}%
  \BibitemOpen
  \bibfield  {author} {\bibinfo {author} {\bibfnamefont {S.}~\bibnamefont
  {Bacca}}, \bibinfo {author} {\bibfnamefont {N.}~\bibnamefont {Barnea}},
  \bibinfo {author} {\bibfnamefont {W.}~\bibnamefont {Leidemann}}, \ and\
  \bibinfo {author} {\bibfnamefont {G.}~\bibnamefont {Orlandini}},\ }\href
  {\doibase 10.1103/PhysRevC.80.064001} {\bibfield  {journal} {\bibinfo
  {journal} {Phys. Rev. C}\ }\textbf {\bibinfo {volume} {80}},\ \bibinfo
  {pages} {064001} (\bibinfo {year} {2009}{\natexlab{b}})}\BibitemShut
  {NoStop}%
\bibitem [{\citenamefont {{A. A. Aguilar-Areval et al.}}(2008)}]{Aguilar:2008}%
  \BibitemOpen
  \bibfield  {author} {\bibinfo {author} {\bibnamefont {{A. A. Aguilar-Areval
  et al.}}} (\bibinfo {collaboration} {MiniBooNE Collaboration}),\ }\href
  {\doibase 10.1103/PhysRevLett.100.032301} {\bibfield  {journal} {\bibinfo
  {journal} {Phys. Rev. Lett.}\ }\textbf {\bibinfo {volume} {100}},\ \bibinfo
  {pages} {032301} (\bibinfo {year} {2008})}\BibitemShut {NoStop}%
\bibitem [{\citenamefont {Benhar}\ \emph {et~al.}(2010)\citenamefont {Benhar},
  \citenamefont {Coletti},\ and\ \citenamefont {Meloni}}]{Benhar:2010}%
  \BibitemOpen
  \bibfield  {author} {\bibinfo {author} {\bibfnamefont {O.}~\bibnamefont
  {Benhar}}, \bibinfo {author} {\bibfnamefont {P.}~\bibnamefont {Coletti}}, \
  and\ \bibinfo {author} {\bibfnamefont {D.}~\bibnamefont {Meloni}},\ }\href
  {\doibase 10.1103/PhysRevLett.105.132301} {\bibfield  {journal} {\bibinfo
  {journal} {Phys. Rev. Lett.}\ }\textbf {\bibinfo {volume} {105}},\ \bibinfo
  {pages} {132301} (\bibinfo {year} {2010})}\BibitemShut {NoStop}%
\bibitem [{\citenamefont {{Amaldi}}\ \emph {et~al.}(1979)\citenamefont
  {{Amaldi}}, \citenamefont {{Fubini}},\ and\ \citenamefont
  {{Furlan}}}]{Amaldi:1979}%
  \BibitemOpen
  \bibfield  {author} {\bibinfo {author} {\bibfnamefont {E.}~\bibnamefont
  {{Amaldi}}}, \bibinfo {author} {\bibfnamefont {S.}~\bibnamefont {{Fubini}}},
  \ and\ \bibinfo {author} {\bibfnamefont {G.}~\bibnamefont {{Furlan}}},\
  }\href {\doibase 10.1007/BFb0048208} {\bibfield  {journal} {\bibinfo
  {journal} {Springer Tracts in Modern Physics}\ }\textbf {\bibinfo {volume}
  {93}} (\bibinfo {year} {1979}),\ 10.1007/BFb0048208}\BibitemShut {NoStop}%
\bibitem [{\citenamefont {Baker}\ \emph {et~al.}(1981)\citenamefont {Baker},
  \citenamefont {Cnops}, \citenamefont {Connolly}, \citenamefont {Kahn},
  \citenamefont {Kirk}, \citenamefont {Murtagh}, \citenamefont {Palmer},
  \citenamefont {Samios},\ and\ \citenamefont {Tanaka}}]{Baker:1981}%
  \BibitemOpen
  \bibfield  {author} {\bibinfo {author} {\bibfnamefont {N.~J.}\ \bibnamefont
  {Baker}}, \bibinfo {author} {\bibfnamefont {A.~M.}\ \bibnamefont {Cnops}},
  \bibinfo {author} {\bibfnamefont {P.~L.}\ \bibnamefont {Connolly}}, \bibinfo
  {author} {\bibfnamefont {S.~A.}\ \bibnamefont {Kahn}}, \bibinfo {author}
  {\bibfnamefont {H.~G.}\ \bibnamefont {Kirk}}, \bibinfo {author}
  {\bibfnamefont {M.~J.}\ \bibnamefont {Murtagh}}, \bibinfo {author}
  {\bibfnamefont {R.~B.}\ \bibnamefont {Palmer}}, \bibinfo {author}
  {\bibfnamefont {N.~P.}\ \bibnamefont {Samios}}, \ and\ \bibinfo {author}
  {\bibfnamefont {M.}~\bibnamefont {Tanaka}},\ }\href {\doibase
  10.1103/PhysRevD.23.2499} {\bibfield  {journal} {\bibinfo  {journal} {Phys.
  Rev. D}\ }\textbf {\bibinfo {volume} {23}},\ \bibinfo {pages} {2499}
  (\bibinfo {year} {1981})}\BibitemShut {NoStop}%
\bibitem [{\citenamefont {{K. L. Miller et al.}}(1982)}]{Miller:1982}%
  \BibitemOpen
  \bibfield  {author} {\bibinfo {author} {\bibnamefont {{K. L. Miller et
  al.}}},\ }\href {\doibase 10.1103/PhysRevD.26.537} {\bibfield  {journal}
  {\bibinfo  {journal} {Phys. Rev. D}\ }\textbf {\bibinfo {volume} {26}},\
  \bibinfo {pages} {537} (\bibinfo {year} {1982})}\BibitemShut {NoStop}%
\bibitem [{\citenamefont {{T. Kitagaki et al.}}(1983)}]{Kitagaki:1983}%
  \BibitemOpen
  \bibfield  {author} {\bibinfo {author} {\bibnamefont {{T. Kitagaki et
  al.}}},\ }\href {\doibase 10.1103/PhysRevD.28.436} {\bibfield  {journal}
  {\bibinfo  {journal} {Phys. Rev. D}\ }\textbf {\bibinfo {volume} {28}},\
  \bibinfo {pages} {436} (\bibinfo {year} {1983})}\BibitemShut {NoStop}%
\bibitem [{\citenamefont {{L. A. Ahrens et al.}}(1987)}]{Ahrens:1987}%
  \BibitemOpen
  \bibfield  {author} {\bibinfo {author} {\bibnamefont {{L. A. Ahrens et
  al.}}},\ }\href {\doibase 10.1103/PhysRevD.35.785} {\bibfield  {journal}
  {\bibinfo  {journal} {Phys. Rev. D}\ }\textbf {\bibinfo {volume} {35}},\
  \bibinfo {pages} {785} (\bibinfo {year} {1987})}\BibitemShut {NoStop}%
\bibitem [{\citenamefont {Carlson}\ \emph {et~al.}(2002)\citenamefont
  {Carlson}, \citenamefont {Jourdan}, \citenamefont {Schiavilla},\ and\
  \citenamefont {Sick}}]{Carlson:2002}%
  \BibitemOpen
  \bibfield  {author} {\bibinfo {author} {\bibfnamefont {J.}~\bibnamefont
  {Carlson}}, \bibinfo {author} {\bibfnamefont {J.}~\bibnamefont {Jourdan}},
  \bibinfo {author} {\bibfnamefont {R.}~\bibnamefont {Schiavilla}}, \ and\
  \bibinfo {author} {\bibfnamefont {I.}~\bibnamefont {Sick}},\ }\href {\doibase
  10.1103/PhysRevC.65.024002} {\bibfield  {journal} {\bibinfo  {journal} {Phys.
  Rev. C}\ }\textbf {\bibinfo {volume} {65}},\ \bibinfo {pages} {024002}
  (\bibinfo {year} {2002})}\BibitemShut {NoStop}%
\bibitem [{\citenamefont {McVoy}\ and\ \citenamefont
  {Van~Hove}(1962)}]{McVoy:1962}%
  \BibitemOpen
  \bibfield  {author} {\bibinfo {author} {\bibfnamefont {K.~W.}\ \bibnamefont
  {McVoy}}\ and\ \bibinfo {author} {\bibfnamefont {L.}~\bibnamefont
  {Van~Hove}},\ }\href {\doibase 10.1103/PhysRev.125.1034} {\bibfield
  {journal} {\bibinfo  {journal} {Phys. Rev.}\ }\textbf {\bibinfo {volume}
  {125}},\ \bibinfo {pages} {1034} (\bibinfo {year} {1962})}\BibitemShut
  {NoStop}%
\bibitem [{\citenamefont {{Jourdan}}(1996)}]{Jourdan:1996}%
  \BibitemOpen
  \bibfield  {author} {\bibinfo {author} {\bibfnamefont {J.}~\bibnamefont
  {{Jourdan}}},\ }\href {\doibase 10.1016/0375-9474(96)00143-1} {\bibfield
  {journal} {\bibinfo  {journal} {Nuclear Physics A}\ }\textbf {\bibinfo
  {volume} {603}},\ \bibinfo {pages} {117} (\bibinfo {year}
  {1996})}\BibitemShut {NoStop}%
\bibitem [{\citenamefont {Schiavilla}\ \emph {et~al.}(1989)\citenamefont
  {Schiavilla}, \citenamefont {Pandharipande},\ and\ \citenamefont
  {Fabrocini}}]{Schiavilla:1989}%
  \BibitemOpen
  \bibfield  {author} {\bibinfo {author} {\bibfnamefont {R.}~\bibnamefont
  {Schiavilla}}, \bibinfo {author} {\bibfnamefont {V.~R.}\ \bibnamefont
  {Pandharipande}}, \ and\ \bibinfo {author} {\bibfnamefont {A.}~\bibnamefont
  {Fabrocini}},\ }\href {\doibase 10.1103/PhysRevC.40.1484} {\bibfield
  {journal} {\bibinfo  {journal} {Phys. Rev. C}\ }\textbf {\bibinfo {volume}
  {40}},\ \bibinfo {pages} {1484} (\bibinfo {year} {1989})}\BibitemShut
  {NoStop}%
\bibitem [{\citenamefont {Schiavilla}\ \emph {et~al.}(1993)\citenamefont
  {Schiavilla}, \citenamefont {Wiringa},\ and\ \citenamefont
  {Carlson}}]{Schiavilla:1993}%
  \BibitemOpen
  \bibfield  {author} {\bibinfo {author} {\bibfnamefont {R.}~\bibnamefont
  {Schiavilla}}, \bibinfo {author} {\bibfnamefont {R.~B.}\ \bibnamefont
  {Wiringa}}, \ and\ \bibinfo {author} {\bibfnamefont {J.}~\bibnamefont
  {Carlson}},\ }\href {\doibase 10.1103/PhysRevLett.70.3856} {\bibfield
  {journal} {\bibinfo  {journal} {Phys. Rev. Lett.}\ }\textbf {\bibinfo
  {volume} {70}},\ \bibinfo {pages} {3856} (\bibinfo {year}
  {1993})}\BibitemShut {NoStop}%
\bibitem [{\citenamefont {Wiringa}\ \emph {et~al.}(1995)\citenamefont
  {Wiringa}, \citenamefont {Stoks},\ and\ \citenamefont
  {Schiavilla}}]{Wiringa:1995}%
  \BibitemOpen
  \bibfield  {author} {\bibinfo {author} {\bibfnamefont {R.~B.}\ \bibnamefont
  {Wiringa}}, \bibinfo {author} {\bibfnamefont {V.~G.~J.}\ \bibnamefont
  {Stoks}}, \ and\ \bibinfo {author} {\bibfnamefont {R.}~\bibnamefont
  {Schiavilla}},\ }\href {\doibase 10.1103/PhysRevC.51.38} {\bibfield
  {journal} {\bibinfo  {journal} {Phys. Rev. C}\ }\textbf {\bibinfo {volume}
  {51}},\ \bibinfo {pages} {38} (\bibinfo {year} {1995})}\BibitemShut {NoStop}%
\bibitem [{\citenamefont {Nollett}\ \emph {et~al.}(2007)\citenamefont
  {Nollett}, \citenamefont {Pieper}, \citenamefont {Wiringa}, \citenamefont
  {Carlson},\ and\ \citenamefont {Hale}}]{Nollett:2007}%
  \BibitemOpen
  \bibfield  {author} {\bibinfo {author} {\bibfnamefont {K.~M.}\ \bibnamefont
  {Nollett}}, \bibinfo {author} {\bibfnamefont {S.~C.}\ \bibnamefont {Pieper}},
  \bibinfo {author} {\bibfnamefont {R.~B.}\ \bibnamefont {Wiringa}}, \bibinfo
  {author} {\bibfnamefont {J.}~\bibnamefont {Carlson}}, \ and\ \bibinfo
  {author} {\bibfnamefont {G.~M.}\ \bibnamefont {Hale}},\ }\href {\doibase
  10.1103/PhysRevLett.99.022502} {\bibfield  {journal} {\bibinfo  {journal}
  {Phys. Rev. Lett.}\ }\textbf {\bibinfo {volume} {99}},\ \bibinfo {pages}
  {022502} (\bibinfo {year} {2007})}\BibitemShut {NoStop}%
\bibitem [{\citenamefont {Pudliner}\ \emph {et~al.}(1997)\citenamefont
  {Pudliner}, \citenamefont {Pandharipande}, \citenamefont {Carlson},
  \citenamefont {Pieper},\ and\ \citenamefont {Wiringa}}]{Pudliner:1997}%
  \BibitemOpen
  \bibfield  {author} {\bibinfo {author} {\bibfnamefont {B.~S.}\ \bibnamefont
  {Pudliner}}, \bibinfo {author} {\bibfnamefont {V.~R.}\ \bibnamefont
  {Pandharipande}}, \bibinfo {author} {\bibfnamefont {J.}~\bibnamefont
  {Carlson}}, \bibinfo {author} {\bibfnamefont {S.~C.}\ \bibnamefont {Pieper}},
  \ and\ \bibinfo {author} {\bibfnamefont {R.~B.}\ \bibnamefont {Wiringa}},\
  }\href {\doibase 10.1103/PhysRevC.56.1720} {\bibfield  {journal} {\bibinfo
  {journal} {Phys. Rev. C}\ }\textbf {\bibinfo {volume} {56}},\ \bibinfo
  {pages} {1720} (\bibinfo {year} {1997})}\BibitemShut {NoStop}%
\bibitem [{\citenamefont {{Sick}}(1982)}]{Sick:1982}%
  \BibitemOpen
  \bibfield  {author} {\bibinfo {author} {\bibfnamefont {I.}~\bibnamefont
  {{Sick}}},\ }\href {\doibase 10.1016/0370-2693(82)90327-6} {\bibfield
  {journal} {\bibinfo  {journal} {Physics Letters B}\ }\textbf {\bibinfo
  {volume} {116}},\ \bibinfo {pages} {212} (\bibinfo {year}
  {1982})}\BibitemShut {NoStop}%
\bibitem [{\citenamefont {Shen}\ \emph {et~al.}(2012)\citenamefont {Shen},
  \citenamefont {Marcucci}, \citenamefont {Carlson}, \citenamefont {Gandolfi},\
  and\ \citenamefont {Schiavilla}}]{Shen:2012}%
  \BibitemOpen
  \bibfield  {author} {\bibinfo {author} {\bibfnamefont {G.}~\bibnamefont
  {Shen}}, \bibinfo {author} {\bibfnamefont {L.~E.}\ \bibnamefont {Marcucci}},
  \bibinfo {author} {\bibfnamefont {J.}~\bibnamefont {Carlson}}, \bibinfo
  {author} {\bibfnamefont {S.}~\bibnamefont {Gandolfi}}, \ and\ \bibinfo
  {author} {\bibfnamefont {R.}~\bibnamefont {Schiavilla}},\ }\href {\doibase
  10.1103/PhysRevC.86.035503} {\bibfield  {journal} {\bibinfo  {journal} {Phys.
  Rev. C}\ }\textbf {\bibinfo {volume} {86}},\ \bibinfo {pages} {035503}
  (\bibinfo {year} {2012})}\BibitemShut {NoStop}%
\bibitem [{\citenamefont {Carlson}\ and\ \citenamefont
  {Schiavilla}(1998)}]{Carlson:1998}%
  \BibitemOpen
  \bibfield  {author} {\bibinfo {author} {\bibfnamefont {J.}~\bibnamefont
  {Carlson}}\ and\ \bibinfo {author} {\bibfnamefont {R.}~\bibnamefont
  {Schiavilla}},\ }\href {\doibase 10.1103/RevModPhys.70.743} {\bibfield
  {journal} {\bibinfo  {journal} {Rev. Mod. Phys.}\ }\textbf {\bibinfo {volume}
  {70}},\ \bibinfo {pages} {743} (\bibinfo {year} {1998})}\BibitemShut
  {NoStop}%
\bibitem [{\citenamefont {Marcucci}\ \emph {et~al.}(2005)\citenamefont
  {Marcucci}, \citenamefont {Viviani}, \citenamefont {Schiavilla},
  \citenamefont {Kievsky},\ and\ \citenamefont {Rosati}}]{Marcucci:2005}%
  \BibitemOpen
  \bibfield  {author} {\bibinfo {author} {\bibfnamefont {L.~E.}\ \bibnamefont
  {Marcucci}}, \bibinfo {author} {\bibfnamefont {M.}~\bibnamefont {Viviani}},
  \bibinfo {author} {\bibfnamefont {R.}~\bibnamefont {Schiavilla}}, \bibinfo
  {author} {\bibfnamefont {A.}~\bibnamefont {Kievsky}}, \ and\ \bibinfo
  {author} {\bibfnamefont {S.}~\bibnamefont {Rosati}},\ }\href {\doibase
  10.1103/PhysRevC.72.014001} {\bibfield  {journal} {\bibinfo  {journal} {Phys.
  Rev. C}\ }\textbf {\bibinfo {volume} {72}},\ \bibinfo {pages} {014001}
  (\bibinfo {year} {2005})}\BibitemShut {NoStop}%
\bibitem [{\citenamefont {Lusk}\ \emph {et~al.}(2010)\citenamefont {Lusk},
  \citenamefont {Pieper},\ and\ \citenamefont {Butler}}]{Lusk:2010}%
  \BibitemOpen
  \bibfield  {author} {\bibinfo {author} {\bibfnamefont {E.}~\bibnamefont
  {Lusk}}, \bibinfo {author} {\bibfnamefont {S.}~\bibnamefont {Pieper}}, \ and\
  \bibinfo {author} {\bibfnamefont {R.}~\bibnamefont {Butler}},\ }\href
  {http://www.scidacreview.org/1002/html/adlb.html} {\bibfield  {journal}
  {\bibinfo  {journal} {SciDAC Review}\ }\textbf {\bibinfo {volume} {17}},\
  \bibinfo {pages} {30} (\bibinfo {year} {2010})},\ \bibinfo {note} {library
  available at http://www.cs.mtsu.edu/$\sim$rbutler/adlb/}\BibitemShut
  {NoStop}%
\bibitem [{\citenamefont {Arvo}(1992)}]{Arvo:1992}%
  \BibitemOpen
  \bibfield  {author} {\bibinfo {author} {\bibfnamefont {J.}~\bibnamefont
  {Arvo}}\ }(\bibinfo  {publisher} {Academic Press Professional, Inc.},\
  \bibinfo {address} {San Diego, CA, USA},\ \bibinfo {year} {1992})\ Chap.\
  \bibinfo {chapter} {Fast random rotation matrices}, pp.\ \bibinfo {pages}
  {117--120}\BibitemShut {NoStop}%
\bibitem [{\citenamefont {{Sick}}(2013)}]{Sick:2013}%
  \BibitemOpen
  \bibfield  {author} {\bibinfo {author} {\bibfnamefont {I.}~\bibnamefont
  {{Sick}}},\ }\href@noop {} {\  (\bibinfo {year} {2013})},\ \bibinfo {note}
  {private communication}\BibitemShut {NoStop}%
\bibitem [{\citenamefont {{Simon}}\ \emph {et~al.}(1980)\citenamefont
  {{Simon}}, \citenamefont {{Schmitt}}, \citenamefont {{Borkowski}},\ and\
  \citenamefont {{Walther}}}]{Simon:1980}%
  \BibitemOpen
  \bibfield  {author} {\bibinfo {author} {\bibfnamefont {G.~G.}\ \bibnamefont
  {{Simon}}}, \bibinfo {author} {\bibfnamefont {C.}~\bibnamefont {{Schmitt}}},
  \bibinfo {author} {\bibfnamefont {F.}~\bibnamefont {{Borkowski}}}, \ and\
  \bibinfo {author} {\bibfnamefont {V.~H.}\ \bibnamefont {{Walther}}},\ }\href
  {\doibase 10.1016/0375-9474(80)90104-9} {\bibfield  {journal} {\bibinfo
  {journal} {Nuclear Physics A}\ }\textbf {\bibinfo {volume} {333}},\ \bibinfo
  {pages} {381} (\bibinfo {year} {1980})}\BibitemShut {NoStop}%
\bibitem [{\citenamefont {{Galster}}\ \emph {et~al.}(1971)\citenamefont
  {{Galster}}, \citenamefont {{Klein}}, \citenamefont {{Moritz}}, \citenamefont
  {{Schmidt}}, \citenamefont {{Wegener}},\ and\ \citenamefont
  {{Bleckwenn}}}]{Galster:1971}%
  \BibitemOpen
  \bibfield  {author} {\bibinfo {author} {\bibfnamefont {S.}~\bibnamefont
  {{Galster}}}, \bibinfo {author} {\bibfnamefont {H.}~\bibnamefont {{Klein}}},
  \bibinfo {author} {\bibfnamefont {J.}~\bibnamefont {{Moritz}}}, \bibinfo
  {author} {\bibfnamefont {K.~H.}\ \bibnamefont {{Schmidt}}}, \bibinfo {author}
  {\bibfnamefont {D.}~\bibnamefont {{Wegener}}}, \ and\ \bibinfo {author}
  {\bibfnamefont {J.}~\bibnamefont {{Bleckwenn}}},\ }\href {\doibase
  10.1016/0550-3213(71)90068-X} {\bibfield  {journal} {\bibinfo  {journal}
  {Nuclear Physics B}\ }\textbf {\bibinfo {volume} {32}},\ \bibinfo {pages}
  {221} (\bibinfo {year} {1971})}\BibitemShut {NoStop}%
\bibitem [{\citenamefont {{H{\"o}hler}}\ \emph {et~al.}(1976)\citenamefont
  {{H{\"o}hler}}, \citenamefont {{Pietarinen}}, \citenamefont
  {{Sabba-Stefanescu}}, \citenamefont {{Borkowski}}, \citenamefont {{Simon}},
  \citenamefont {{Walther}},\ and\ \citenamefont {{Wendling}}}]{Hohler:1976}%
  \BibitemOpen
  \bibfield  {author} {\bibinfo {author} {\bibfnamefont {G.}~\bibnamefont
  {{H{\"o}hler}}}, \bibinfo {author} {\bibfnamefont {E.}~\bibnamefont
  {{Pietarinen}}}, \bibinfo {author} {\bibfnamefont {I.}~\bibnamefont
  {{Sabba-Stefanescu}}}, \bibinfo {author} {\bibfnamefont {F.}~\bibnamefont
  {{Borkowski}}}, \bibinfo {author} {\bibfnamefont {G.~G.}\ \bibnamefont
  {{Simon}}}, \bibinfo {author} {\bibfnamefont {V.~H.}\ \bibnamefont
  {{Walther}}}, \ and\ \bibinfo {author} {\bibfnamefont {R.~D.}\ \bibnamefont
  {{Wendling}}},\ }\href {\doibase 10.1016/0550-3213(76)90449-1} {\bibfield
  {journal} {\bibinfo  {journal} {Nuclear Physics B}\ }\textbf {\bibinfo
  {volume} {114}},\ \bibinfo {pages} {505} (\bibinfo {year}
  {1976})}\BibitemShut {NoStop}%
\bibitem [{\citenamefont {de~Vries}\ \emph {et~al.}(1987)\citenamefont
  {de~Vries}, \citenamefont {de~Jager},\ and\ \citenamefont
  {de~Vries}}]{deVries:1987}%
  \BibitemOpen
  \bibfield  {author} {\bibinfo {author} {\bibfnamefont {H.}~\bibnamefont
  {de~Vries}}, \bibinfo {author} {\bibfnamefont {E.~E.}\ \bibnamefont
  {de~Jager}}, \ and\ \bibinfo {author} {\bibfnamefont {C.}~\bibnamefont
  {de~Vries}},\ }\href@noop {} {\bibfield  {journal} {\bibinfo  {journal} {At.
  Data Nucl. Data Tables}\ }\textbf {\bibinfo {volume} {36}},\ \bibinfo {pages}
  {495} (\bibinfo {year} {1987})}\BibitemShut {NoStop}%
\bibitem [{\citenamefont {Entem}\ and\ \citenamefont
  {Machleidt}(2003)}]{Entem:2003}%
  \BibitemOpen
  \bibfield  {author} {\bibinfo {author} {\bibfnamefont {D.}~\bibnamefont
  {Entem}}\ and\ \bibinfo {author} {\bibfnamefont {R.}~\bibnamefont
  {Machleidt}},\ }\href {\doibase 10.1103/PhysRevC.68.041001} {\bibfield
  {journal} {\bibinfo  {journal} {Phys.Rev.}\ }\textbf {\bibinfo {volume}
  {C68}},\ \bibinfo {pages} {041001} (\bibinfo {year} {2003})},\ \Eprint
  {http://arxiv.org/abs/nucl-th/0304018} {arXiv:nucl-th/0304018 [nucl-th]}
  \BibitemShut {NoStop}%
\bibitem [{\citenamefont {Lu}\ \emph {et~al.}(1999)\citenamefont {Lu},
  \citenamefont {Tsushima}, \citenamefont {Thomas}, \citenamefont {Williams},\
  and\ \citenamefont {Saito}}]{Lu:1999}%
  \BibitemOpen
  \bibfield  {author} {\bibinfo {author} {\bibfnamefont {D.~H.}\ \bibnamefont
  {Lu}}, \bibinfo {author} {\bibfnamefont {K.}~\bibnamefont {Tsushima}},
  \bibinfo {author} {\bibfnamefont {A.~W.}\ \bibnamefont {Thomas}}, \bibinfo
  {author} {\bibfnamefont {A.~G.}\ \bibnamefont {Williams}}, \ and\ \bibinfo
  {author} {\bibfnamefont {K.}~\bibnamefont {Saito}},\ }\href {\doibase
  10.1103/PhysRevC.60.068201} {\bibfield  {journal} {\bibinfo  {journal} {Phys.
  Rev. C}\ }\textbf {\bibinfo {volume} {60}},\ \bibinfo {pages} {068201}
  (\bibinfo {year} {1999})}\BibitemShut {NoStop}%
\bibitem [{\citenamefont {Guichon}\ and\ \citenamefont
  {Thomas}(2004)}]{Guichon:2004}%
  \BibitemOpen
  \bibfield  {author} {\bibinfo {author} {\bibfnamefont {P.~A.~M.}\
  \bibnamefont {Guichon}}\ and\ \bibinfo {author} {\bibfnamefont {A.~W.}\
  \bibnamefont {Thomas}},\ }\href {\doibase 10.1103/PhysRevLett.93.132502}
  {\bibfield  {journal} {\bibinfo  {journal} {Phys. Rev. Lett.}\ }\textbf
  {\bibinfo {volume} {93}},\ \bibinfo {pages} {132502} (\bibinfo {year}
  {2004})}\BibitemShut {NoStop}%
\end{thebibliography}%

\end{document}